\documentclass[12pt]{iopart}
\usepackage{iopams}
\usepackage{setstack}

\usepackage{amstext}
\usepackage{enumerate}
\usepackage{url}
\usepackage{epsf}
\usepackage{bm}

\newenvironment{proof}{\noindent Proof.\quad}{}
\newcommand{\binom}[2]{{\begingroup#1\endgroup\over#2}}
\newcommand{\lvert}{|}
\newcommand{\rvert}{|}

\newcommand{\I}{I}

\begin{document}

\title[Two-party Bell inequalities via triangular elimination]
  {Two-party Bell inequalities derived from combinatorics via
   triangular elimination}

\author{David Avis$\dag$, Hiroshi Imai$\ddag\S$,
  Tsuyoshi Ito$\ddag$ and Yuuya Sasaki$\ddag$}

\address{\dag\ School of Computer Science, McGill University,
         3480 University, Montreal, Quebec, Canada H3A 2A7}

\address{\ddag\ Department of Computer Science, University of Tokyo,
         7-3-1 Hongo, Bunkyo-ku, Tokyo 113-0033, Japan}

\address{\S\ ERATO Quantum Computation and Information Project,
         5-28-3 Hongo, Bunkyo-ku, Tokyo 113-0033, Japan}

\eads{\mailto{avis@cs.mcgill.ca},
  \mailto{\symbol{`\{}imai,tsuyoshi,y\_sasaki\symbol{`\}}@is.s.u-tokyo.ac.jp}}

\begin{abstract}
  We establish a relation between the two-party Bell
  inequalities for two-valued measurements and a high-dimensional
  convex polytope called the cut polytope in polyhedral combinatorics.
  Using this relation, we propose a method, \emph{triangular
  elimination}, to derive tight Bell inequalities from facets of the cut
  polytope.
  This method gives two hundred million inequivalent tight Bell
  inequalities from
  currently known results on the cut polytope.
  In addition, this method gives general formulas which represent
  families of infinitely many Bell inequalities.
  These results can be used to examine general properties of Bell
  inequalities.
\end{abstract}



\maketitle



\newcommand{\vct}{\bm}

\newcommand{\RR}{{\mathbb{R}}}
\newcommand{\A}{{\mathrm{A}}}
\newcommand{\B}{{\mathrm{B}}}
\newcommand{\X}{{\mathrm{X}}}
\newcommand{\K}{{\mathrm{K}}}
\newcommand{\trans}{{\mathrm{T}}}
\newcommand{\CutP}{\mathrm{CUT}^\square}
\newcommand{\CutC}{\mathrm{CUT}}
\newcommand{\prob}{\mathop{\mbox{Pr}}}

\makeatletter
\newcommand{\revddots}
  {\mathinner{\mkern1mu\raise\p@
   \vbox{\kern7\p@\hbox{.}}\mkern2mu
   \raise4\p@\hbox{.}\mkern2mu\raise7\p@\hbox{.}\mkern1mu}}
\makeatother


\newtheorem{theorem}{Theorem}[section]
\newtheorem{lemma}[theorem]{Lemma}
\newtheorem{corollary}[theorem]{Corollary}
\newtheorem{proposition}[theorem]{Proposition}
\newtheorem{observation}[theorem]{Observation}
\newtheorem{conjecture}[theorem]{Conjecture}
\newtheorem{fact}[theorem]{Fact}
\newtheorem{claim}{Claim}[section]

\newtheorem{definition}{Definition}[section]
\newtheorem{example}{Example}[section]

\newtheorem{remark}{Remark}[section]



\section{Introduction}

Bell inequalities have been intensively studied in quantum
theory~\cite{WerWol-QIC01,KruWer-0504166}, and it is known that 
they can be obtained from the structure
of certain convex
polytopes~\cite{Pit-JMP86,Pit-MP91,Per:all99}.
Bell inequalities are not the only example of the use of convex polytopes
in quantum theory.
In a pioneering paper, McRae and Davidson~\cite{McrDav-JMP72} used 
the theory of convex polytopes to
obtain inequalities bounding the range of possible solutions to some
problems in quantum mechanics.
Their method is summarized as follows:
First they prove that the possible solutions form a convex polytope
and obtain the set of vertices of the polytope.
Then they obtain a minimum set of inequalities that describe the polytope
using a convex hull algorithm.
Interestingly, one of the polytopes McRae and Davidson considered
coincides with the \emph{correlation polytope} Pitowsky introduced in
\cite{Pit-JMP86}, in connection with Bell inequalities.
This polytope arises in many fields under different names, and a
comprehensive source for results on this polytope and the related cut
polytope (described later) is the book by Deza and Laurent~\cite{DezLau:cut97}.

In this paper we consider the results of correlation experiments
between two parties, where one party has 
$m_\A$ choices of possible two-valued measurements
and the other party has $m_\B$ choices.
The relevant polytope can be
described as follows.
The results of a series of such correlation experiments
are represented as a
vector of $m_\A+m_\B+m_\A m_\B$ probabilities.
In classical mechanics, the set of vectors which are possible
results of a correlation experiment forms an
$(m_\A+m_\B+m_\A m_\B)$-dimensional convex polytope which
is a projection of the correlation polytope onto the complete
bipartite graph $K_{m_\A , m_\B}$.
A Bell inequality is nothing but a linear inequality 
satisfied by all the points in such a polytope. 
A tight Bell inequality is a Bell inequality which cannot be
represented as a positive weighted sum of other Bell inequalities 
and defines a
facet of the polytope. Two examples of these facet defining inequalities 
are the nonnegativity 
inequality and the Clauser-Horne-Shimony-Holt (CHSH)
inequality~\cite{ClaHorShiHol-PRL69}.

By considering these polytopes, Bell's original
inequality~\cite{Bel-Phys64}, the CHSH
and many other known Bell inequalities
can be understood in a unified
manner. 
Fine's necessary and sufficient conditions~\cite{Fin-PRL82} for
$m_\A=m_\B=2$ can be seen as the complete inequality representation of
the correlation polytope of the complete bipartite graph $\K_{2,2}$.
Pitowsky and Svozil~\cite{PitSvo-PRA01} and Collins and
Gisin~\cite{ColGis-JPA04} apply convex hull algorithms to obtain a 
complete list of tight Bell inequalities in other experimental settings.
As a result, we know the complete list of Bell inequalities in the
cases $m_\A=2$~\cite{ColGis-JPA04},
$(m_\A,m_\B)=(3,3)$~\cite{PitSvo-PRA01} and
$(m_\A,m_\B)=(3,4)$~\cite{ColGis-JPA04}.
Several
software packages for convex hull computation such as cdd~\cite{Fuk:cdd} and lrs~\cite{Avi:lrs}
are readily available.
It is unlikely, however,
that there exists a compact representation of the complete set of Bell
inequalities in arbitrarily large settings.
This follows from the fact that testing whether a vector of correlations
lies in the correlation polytope of the bipartite graph $K_{m_\A,m_\B}$
is NP-complete~\cite{AviImaItoSas:0404014}.
Therefore it is natural to look for families of Bell inequalities,
especially those that are facet producing. 
In this direction, Collins and Gisin~\cite{ColGis-JPA04} give a family
$\I_{mm22}$ of Bell inequalities in the case $m_\A=m_\B=m$ for
general $m$.
In addition there are several
extensions~\cite{ColGisLinMasPop-PRL02,Mas-QIC03,ColGis-JPA04} of the
CHSH inequality for multi-valued measurements.

In the field of polyhedral combinatorics a polytope isomorphic to
the correlation polytope, called the \emph{cut polytope},
has been studied in great detail~\cite{DezLau:cut97}.
The correlation and cut polytopes
are isomorphic via a linear mapping~\cite{Ham-OR65}
and so the inequalities representing them correspond one-to-one.
This relationship enables us to apply results for the cut polytope to
the study of
Bell inequalities.
Related to this, Pironio~\cite{Pir-JMP05} uses lifting, which is a
common approach in combinatorial optimization, to generate tight Bell
inequalities for a larger system from those for a smaller system.
Since the mathematical description of the facet structure of cut
polytopes is simpler than that for correlation polytopes, the former
are preferred in polyhedral combinatorics.
Large classes of facets for the cut polytope $\CutP_n$ of the complete
graph $\K_n$ are known for general $n$ \cite{DezLau:cut97}, 
and a complete or conjectured
complete list of all facets for $\CutP_n$ is known for
$n\le9$~\cite{SMAPO}.
We make use of these results in this paper.

The cut polytope of the complete graph has been
the most extensively studied.
However, the case we are interested in corresponds to the correlation
polytope of the complete bipartite graph $\K_{m_\A,m_\B}$, which maps
to the cut polytope of the complete tripartite graph
$\K_{1,m_\A,m_\B}$.
To overcome this gap, we introduce a method called triangular
elimination to convert an inequality valid for $\CutP_n$ to another
inequality valid for $\CutP(\K_{1,m_\A,m_\B})$, which is then
converted to a Bell inequality via the isomorphism.
The CHSH inequality and some of the other previously known
inequalities can be explained in this manner.
More importantly, triangular elimination converts a facet inequality
of $\CutP_n$ to a facet inequality of $\CutP(\K_{1,m_\A,m_\B})$, which
corresponds to a tight Bell inequality.

A complete list of facets of $\CutP_n$ for $n \leq 7$ 
and a conjectured complete list for $n=8,9$ are known.
We apply triangular elimination to these facets to obtain
201,374,783 tight Bell inequalities.
On the other hand, several formulas which represent many different
inequalities valid for $\CutP_n$ are known.
We apply triangular elimination to these formulas to obtain new families
of Bell inequalities.
We discuss their properties such as tightness and inclusion of the
CHSH inequality.

The rest of this paper is organized as follows.
In Section~\ref{sect:elimination}, we introduce
triangular elimination to derive tight Bell inequalities from facets
of the cut polytope of the complete graph, and show its properties.
We also give a computational result on the
number of Bell inequalities obtained by triangular elimination.
In Section~\ref{sect:families} we apply triangular elimination to some
of the known classes of facets of the cut polytope of the complete
graph to obtain general formulas representing many Bell inequalities.
Section~\ref{sect:concluding} concludes the paper by giving the
relation of our result to some of the open problems posed
in~\cite{KruWer-0504166}.

\section{Triangular elimination} \label{sect:elimination}

   \subsection{Bell inequalities and facets of cut polytopes}

Consider a system composed of subsystems $\A$ (Alice) and $\B$ (Bob).
Suppose that on both subsystems, one of $m_\A$ observables for Alice and 
one of $m_\B$ observables for Bob are measured.
For each observable, the outcome
is one of two values (in the rest of the paper, we label the outcomes as $0$ or $1$).
The experiment is repeated a large number of times.
The result of such a correlation experiment consists of the probability distribution of 
the $m_\A m_\B$ joint measurements by both parties.
Throughout this paper, we represent the experimental result as a vector $\vct{q}$
in $m_\A + m_\B + m_\A m_\B$ dimensional space in the following manner:
$q_{A_{i}}$, $q_{B_{j}}$ and
$q_{A_{i}B_{j}}$ correspond to the probabilities  $\prob[A_{i} = 1]$, $\prob[B_{j} = 1]$
and $\prob[A_{i} = 1 \wedge B_{j} = 1]$
 respectively.

In classical mechanics, the result of a correlation experiment must correspond to a probability distribution over 
all \emph{classical configurations}, where a classical configuration
is an assignment of the outcomes $\{0,1\}$
to each of the $m_\A + m_\B$ observables. The experimental result
has a \emph{local hidden variable model}
if and only if a given experimental result can be interpreted as a 
result of such a classical correlation experiment.

\emph{Bell inequalities} are valid linear inequalities for every
experimental result which has a local hidden variable model.
Specifically using the above formulation, we represent a Bell inequality in the form 
\[ \sum_{ 1 \leq i \leq m_\A}b_{A_{i}}q_{A_{i}} + \sum_{ 1 \leq j \leq m_\B}b_{B_{j}}q_{B_{j}} +  \sum_{ 1 \leq i \leq m_\A,  1 \leq j \leq m_\B}b_{A_{i}B_{j}}q_{A_{i}B_{j}} \leq b_{0}. \]
for suitably chosen constants $b_x$.

For example, Clauser, Horn, Shimony and Holt~\cite{ClaHorShiHol-PRL69} 
have shown that the following CHSH inequality is a valid Bell inequality:
\[ -q_{A_{1}} -q_{B_{1}} + q_{A_{1}B_{1}} + q_{A_{1}B_{2}} + q_{A_{2}B_{1}} - q_{A_{2}B_{2}} \leq 0. \]

In general, the set of all experimental results with a local hidden variable model
forms a convex polytope with extreme points corresponding to the classical configurations.
If the results of the experiment are in the above form, the polytope
is called a \emph{correlation polytope}, a name introduced by
Pitowsky~\cite{Pit:prob89}.
(Such polyhedra have been discovered and rediscovered several times, see for instance Deza and Laurent~\cite{DezLau:cut97}.) 
From such a viewpoint, Bell inequalities can be considered as the boundary, or face inequalities, of that polytope.
Since every polytope is the intersection of finitely many half spaces represented by linear inequalities, 
every Bell inequality can be represented by a convex combination of finitely many extremal inequalities. 
Such extremal inequalities are called \emph{tight} Bell
inequalities. Non-extremal inequalities are called \emph{redundant}.

In polytopal theory, the maximal extremal faces of a polytope are
called \emph{facets}.
Therefore, tight Bell inequalities are facet inequalities of the polytope formed by
experimental results with a local hidden variable model.
Note that for a given linear inequality $\vct{b}^{\trans} \vct{q} \leq  b_{0}$ and $d$ dimensional polytope,
the face
represented by the inequality is a facet of that polytope if and only if the dimension of the convex hull of
the extreme points for which the equality holds is $d-1$.

\subsubsection{Cut polytope of complete tripartite graph}
We introduce a simple representation of an experimental setting as a graph.
Consider a graph which consists of vertices corresponding to observables $A_{i}$ or $B_{j}$ and
edges corresponding to joint measurements between  $A_{i}$ and $B_{j}$.
In addition, to represent
probabilities which are the results of single (not joint) measurements,
we introduce a vertex $X$ (which represents the trace out operation of
the other party) and edges between $X$ and $A_{i}$ for every
 $1 \leq i \leq m_\A$, and between $X$ and $B_{j}$ for every
$1 \leq j \leq m_\B$.
This graph is a complete
 tripartite graph since there exist edges between each party of vertices (observables) $\{ X\}$,
 $\{ A_{i}\}$ and $\{ B_{j}\}$. Using this graph, we can conveniently represent either the result probabilities or the coefficients of 
a Bell inequality as edge labels.
We denote this graph by $\K_{1,m_\A,m_\B}$.

In polyhedral combinatorics, a polytope affinely isomorphic to the correlation polytope
has been well studied.
Specifically, if we consider the probabilities $x_{A_{i}B_{j}} = \prob[A_{i} \neq B_{j}]$ instead of $q_{A_{i}B_{j}} = \prob[A_{i} = 1 \wedge B_{j} = 1]$ for each edge,
the probabilities form a polytope called the \emph{cut polytope}.
Thus, the cut polytope
is another formulation of the polytope formed by Bell inequalities.

A \emph{cut} in a graph is an assignment of $\{0,1\}$ to each vertex,
$1$ to an edge between vertices with different
values assigned, and $0$ to an edge between vertices with the same values assigned.
In the above formulation, each cut corresponds to a classical
configuration. Note that since the $0,1$ exchange of all values of vertices yields
the  same edge cut, we can without loss of generality
assume that the vertex $X$ is always assigned the label $0$. 

Let the \emph{cut vector} $\vct{\delta}'(S') \in \RR^{ \{XA_{i}\} \cup \{XB_{j}\} \cup \{A_{i}B_{j}\} }$ for some cut $S'$ be
$\delta'_{uv}(S') = 1$ if vertices $u$ and $v$ are assigned different values, and $0$ if assigned the 
same values. Then, the convex combination
of all the cut vectors $\CutP(\K_{1,m_\A,m_\B}) = \left\{ \vct{x} = \sum_{S'\colon\text{cut}} \lambda_{S'}\vct{\delta}'(S')\mid
 \sum_{S'\colon\text{cut}} \lambda_{S'} = 1 \text{ and } \lambda_{S'} \geq 0   \right\}$ is called the
cut polytope of the
 complete tripartite graph.
The cut polytope has full dimension. Therefore, $\dim(\CutP(\K_{1,m_\A,m_\B})) = m_\A + m_\B + m_\A m_\B$.

In this formulation, a tight Bell inequality $\vct{b}^{\trans} \vct{q} \leq b_{0}$ corresponds to a facet inequality
$\vct{a}^{\prime\trans} \vct{x} \leq a_{0}$ of the cut polytope. The affine isomorphisms between them
are:
\begin{equation}
  \fl
    \left\{ \begin{array}{l}
    x_{XA_{i}}=q_{A_{i}},\\
    x_{XB_{j}}=q_{B_{j}},\\
    x_{A_{i}B_{j}}= q_{A_{i}} + q_{B_{j}} - 2q_{A_{i}B_{j}},
    \end{array} \right. \qquad \left\{
    \begin{array}{l}
    q_{A_{i}}=x_{XA_{i}},\\
    q_{B_{j}}=x_{XB_{j}},\\
    q_{A_{i}B_{j}}= \frac{1}{2}(x_{XA_{i}} + x_{XB_{j}} - x_{A_{i}B_{j}}).
    \end{array} \right.
  \label{eq:isomorphism}
\end{equation}

Actually, because
cut polytopes are symmetric under the switching operation (explained
in Section~\ref{symmetry-of-cut-polytope}) we can
assume that the right hand side of a facet inequality of the cut polytope is always $0$. This means that a given Bell inequality
is tight if and only if for the corresponding facet inequality $\vct{a}^{\trans} \vct{x} \leq 0$ of the cut polytope,
there exist $m_\A + m_\B + m_\A m_\B - 1$ linearly independent cut vectors $\vct{\delta}'(S')$ for which
$\vct{a}^{\prime\trans} \vct{\delta}'(S') = 0$.

For example, there exists a facet inequality $-x_{A_{1}B_{1}} -x_{A_{1}B_{2}} -x_{A_{2}B_{1}} +x_{A_{2}B_{2}} \leq 0$
for $\CutP(\K_{1,m_\A, m_\B})$, $1 \leq m_\A, m_\B$ which corresponds to the CHSH inequality. Therefore, the CHSH inequality is
tight in addition to being valid.

A consequence of the above affine isomorphisms is that any theorem concerning facets of the cut polytope
can be immediately translated to give a corresponding theorem for tight Bell inequalities.
Recently, Collins and Gisin~\cite{ColGis-JPA04} gave the following conjecture about the tightness of Bell inequalities:
if a Bell inequality $\vct{b}^{\trans} \vct{q} \leq b_{0}$ is tight in a given setting $m_\A, m_\B$, 
then for each $ m^{\prime}_{A} \geq m_\A$ and $m^{\prime}_{B} \geq m_\B$,
the inequality $\vct{b}^{\prime\trans} \vct{q}^{\prime} \leq b_{0}$ is also tight. Here $\vct{b}^{\prime}$ is the vector
$b_{uv}^{\prime} = b_{uv}$ if the vertices (observables) $u,v$ appear in $\vct{b}$ and is zero otherwise. 
They gave empirical evidence for this conjecture based on
numerical experiments. In fact, a special case of the \emph{zero-lifting theorem} by De~Simone~\cite{Des-ORL90}
gives a proof of their conjecture. 

  \subsection{Triangular elimination}
      \subsubsection{Cut polytope of complete graph}
In the previous section we saw that the problem of enumerating tight Bell inequalities
is equivalent to that of enumerating facet inequalities of the cut polytope of a corresponding 
complete tripartite graph.
The properties of facet inequalities of
the cut polytope of the complete graph $\K_n$ are well studied and there are rich results.
For example, several general classes of facet inequalities with
relatively simple representations are known.

For $n \le 7$ the complete list of facets is known~\cite{Gri-EJC90}, and
for $n=8,9$ a conjectured complete list is known~\cite{ChrRei-IJCGA01,SMAPO}.
In addition, the symmetry of the
polytope is also well-understood.
We show how to apply such results to our complete tripartite graph case.

First, we introduce the cut polytope of complete graph. The graph is denoted by $\K_n$, has $n$
vertices, and has an edge between each pair of vertices.
As before, a cut is an
assignment of $\{ 0,1\}$ to each vertex, and an edge is labeled by $1$ if
the endpoints of the edge are labeled differently or $0$ if labeled the same. The cut vectors
${\delta}(S)$ of the complete graph are defined in the same manner as before. The
set of all convex combinations of cut vectors $\CutP(\K_{n}) = \left\{ \vct{x} = \sum_{S\colon\text{cut}} \lambda_{S} \vct{\delta}(S)\mid
\sum_{S\colon\text{cut}} \lambda_{S} = 1 \text{ and } \lambda_{S} \geq 0\right\}$
is called the cut polytope of the complete graph.
$\CutP(\K_n)$ is also written as $\CutP_n$.

In contrast to the complete tripartite graph, the space on which the
cut polytope of the complete graph exists has elements corresponding to
probabilities of joint measurement by the same party. Because of the
restrictions of quantum mechanics, such joint measurements are prohibited.
Therefore, if we want to generate tight Bell inequalities from the known
facet inequalities of the cut polytope of the complete graph, we must
transform the inequalities to eliminate joint measurement terms.
In polyhedral terms, $\CutP(\K_{1,m_\A, m_\B})$ is a projection of 
$\CutP(\K_{n})$ onto a lower dimensional space.

     \subsubsection{Definition of triangular elimination}

\begin{figure}
  \begin{center}
    \raisebox{-0.5\height}{\epsfbox{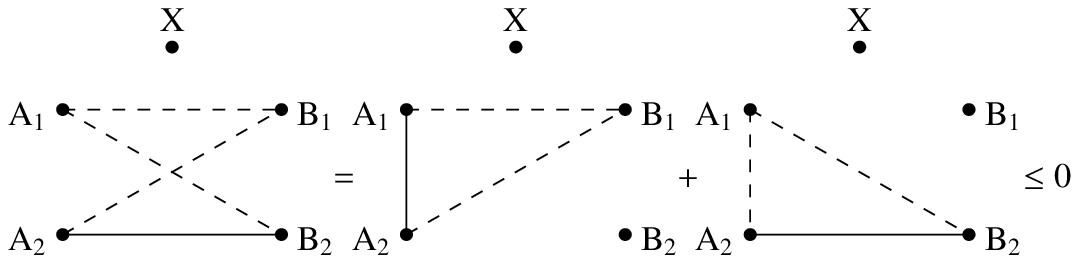}}
    \quad
    \raisebox{-0.5\height}{\epsfbox{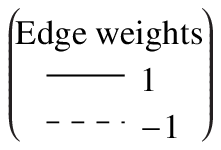}}
  \end{center}
  \caption{\label{fig:chsh}
    The most simple case of triangular elimination:
    The sum of two triangle inequalities is the CHSH inequality.}
\end{figure}

A well known method for projecting a polytope
is called Fourier-Motzkin elimination.
This is essentially the summation of two
facet inequalities to cancel out the target term. For example,
it is well known that the triangle inequality $x_{uv} - x_{uw} - x_{wv} \leq 0$, for any
three vertices $u,v,w$, is valid for the cut polytope of the complete
graph.
In fact, Bell's original inequality~\cite{Bel-Phys64} is essentially this inequality.
The CHSH inequality $-x_{A_{1}B_{1}} -x_{A_{1}B_{2}} -x_{A_{2}B_{1}}
+x_{A_{2}B_{2}} \leq 0$ is the sum of
$x_{A_{1}A_{2}} - x_{A_{1}B_{1}} - x_{A_{2}B_{1}} \leq 0$ and
$x_{A_{2}B_{2}}  -x_{A_{1}B_{2}} - x_{A_{1}A_{2}} \leq 0$
(see \fref{fig:chsh}).

In general, the result of Fourier-Motzkin elimination is not necessarily a facet.
For example, 
it is known that the pentagonal inequality 
\begin{equation}
  \fl
  x_{XA_{1}} + x_{XA_{2}} - x_{XB_{1}} - x_{XB_{2}} + x_{A_{1}A_{2}} - x_{A_{1}B_{1}} - x_{A_{1}B_{2}} - x_{A_{2}B_{1}} - x_{A_{2}B_{2}} + x_{B_{1}B_{2}} \leq 0
  \label{eq:pent}
\end{equation}
is a facet inequality of $\CutP(\K_{5})$. 
If we eliminate joint measurement terms
$x_{A_{1}A_{2}}$ and  $x_{B_{1}B_{2}}$ by adding triangle inequalities 
$x_{A_{1}B_{2}} - x_{A_{1}A_{2}} - x_{A_{2}B_{2}} \leq 0$ and
$x_{A_{2}B_{1}} - x_{B_{1}B_{2}} - x_{A_{2}B_{2}} \leq 0$, the result is 
$x_{XA_{1}} + x_{XA_{2}} - x_{XB_{1}} - x_{XB_{2}} - x_{A_{1}B_{1}} - 3x_{A_{2}B_{2}} \leq 0$. 
Therefore, this inequality is a valid inequality for
$\CutP(\K_{1,3,3})$.
However, the inequality is a summation of four valid triangle inequalities for $\CutP(\K_{1,3,3})$,
namely $x_{XA_{1}} - x_{XB_{1}} - x_{A_{1}B_{1}} \leq 0$, $x_{XA_{2}} - x_{XB_{2}} - x_{A_{2}B_{2}} \leq 0$,
$x_{XA_{2}} - x_{XB_{2}} - x_{A_{2}B_{2}} \leq 0$ and $- x_{XA_{2}} + x_{XB_{2}} - x_{A_{2}B_{2}} \leq 0$.
This means that the inequality with eliminated terms is redundant.

Fourier-Motzkin elimination often produces large numbers of redundant inequalities,
causing the algorithm to be computationally intractable when iterated many times.
Therefore, it is important to find situations where the new inequalities found
are guaranteed to be tight.

The difference between the two examples is that in the CHSH case, the second triangle inequality
introduced a new vertex $B_{2}$ where ``new'' means that the first triangle inequality had no term
with subscript labeled ${B_{2}}$. Generalizing this operation, we will show that 
Fourier-Motzkin elimination by triangle inequalities which introduce new vertices, is 
almost always guaranteed to yield non-redundant inequalities.
We call the operation \emph{triangular elimination}.

\begin{definition}[triangular elimination]

For a given valid inequality for $\CutP(\K_{1 + n_\A + n_\B})$
\begin{eqnarray}
\sum_{ 1 \leq i \leq n_\A}a_{XA_{i}}x_{XA_{i}} + \sum_{ 1 \leq j \leq n_\B}a_{XB_{j}}x_{XB_{j}} + \sum_{ 1 \leq i \leq n_\A,  1 \leq j \leq n_\B}a_{A_{i}B_{j}}x_{A_{i}B_{j}} \nonumber\\
+\sum_{ 1 \leq i < i^{\prime} \leq n_\A}a_{A_{i}A_{i^{\prime}}}x_{A_{i}A_{i^{\prime}}}
+\sum_{ 1 \leq j < j^{\prime} \leq n_\B}a_{B_{j}B_{j^{\prime}}}x_{B_{j}B_{j^{\prime}}}
 \leq a_{0}, \label{eq:before-elimination}
\end{eqnarray}
the triangular elimination is defined as follows:
\begin{eqnarray}
    \sum_{ 1 \leq i \leq n_\A}a_{XA_{i}}x_{XA_{i}}
  + \sum_{ 1 \leq j \leq n_\B}a_{XB_{j}}x_{XB_{j}}
  + \sum_{ 1 \leq i \leq n_\A,  1 \leq j \leq n_\B}
      a_{A_{i}B_{j}}x_{A_{i}B_{j}} \nonumber\\
  + \sum_{ 1 \leq i < i^{\prime} \leq n_\A}
      ( a_{A_{i}A_{i^{\prime}}}x_{A_{i}B'_{A_{i}A_{i^{\prime}}}}
       - |a_{A_{i}A_{i^{\prime}}}| x_{A_{i^{\prime}}B'_{A_{i}A_{i^{\prime}}}}
      ) \nonumber\\
  +\sum_{ 1 \leq j < j^{\prime} \leq n_\B}
      ( a_{B_{j}B_{j^{\prime}}}x_{A'_{B_{j}B_{j^{\prime}}}B_{j}}
       - |a_{A_{j}A_{j^{\prime}}}|x_{A'_{B_{j}B_{j^{\prime}}}B_{j^{\prime}}}
      )
   \leq a_{0}. \label{eq:after-elimination}
\end{eqnarray}
This is an inequality for $\CutP(\K_{1,m_\A,m_\B})$,
where $m_\A = n_\A + \frac{n_\B(n_\B-1)}{2},m_\B = n_\B + \frac{n_\A(n_\A-1)}{2}$.
We denote (\ref{eq:before-elimination}) by $\vct{a}^{\trans}\vct{x}
\leq 0$, $\vct{a},\vct{x}
\in\RR^{\frac{(n_\A + n_\B)(n_\A + n_\B + 1)}{2}} $ and
(\ref{eq:after-elimination})
by $\vct{a}^{\prime\trans}\vct{x}^{\prime} \leq 0, \vct{a}^{\prime},\vct{x}^{\prime} \in\RR^{m_\A + m_\B + m_\A m_\B}$, respectively.
\end{definition}

Note that forbidden terms of the form $x_{A_{i}A_{i^{\prime}}}$ and $x_{B_{j}B_{j^{\prime}}}$
do not appear
in (\ref{eq:after-elimination}).

\begin{figure}
  \begin{center}
    \epsfbox{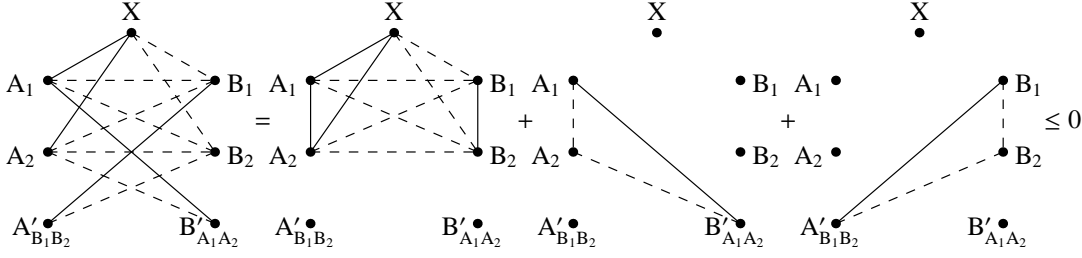}
  \end{center}
  \caption{\label{fig:i3322}
  The $\I_{3322}$ inequality is generated by triangular
  elimination from the pentagonal inequality of $\CutP_5$.}
\end{figure}

As an example, let us see how the $\I_{3322}$ inequalities is
generated by triangular elimination (see \fref{fig:i3322})
of the pentagonal inequality (\ref{eq:pent}).
This inequality has two terms $x_{\A_1\A_2}$ and $x_{\B_1\B_2}$ which
correspond to joint measurements of two observables in one subsystem
and are not allowed.
Therefore, we eliminate these terms by adding two new nodes
$\A'_{\B_1\B_2}$ and $\B'_{\A_1\A_2}$ and adding two triangle
inequalities
$-x_{\A_1\A_2}+x_{\A_1\B'_{\A_1\A_2}}-x_{\A_2\B'_{\A_1\A_2}}\le0$ and
$-x_{\B_1\B_2}+x_{\A'_{\B_1\B_2}\B_1}-x_{\A'_{\B_1\B_2}\B_2}\le0$.
If we rewrite the resulting inequality in terms of the vector
$\vct{q}$ instead of the vector $\vct{x}$ by using the
isomorphism~(\ref{eq:isomorphism}), this inequality becomes the
$\I_{3322}$ inequality.
As we will see in the next subsection, this gives another proof of the
tightness of the $\I_{3322}$ inequality than directly checking the
dimension of the face computationally.

   \subsection{Triangular elimination and facet}
In this subsection, we show the main theorem of this paper: under a very mild condition, the 
triangular elimination of a facet is a facet.

\begin{theorem} \label{thrm:kn-k1mm}
  The triangular elimination of a facet inequality
  $\vct{a}^{\trans}\vct{x}\le0$ of $\CutP(\K_{1+n_\A+n_\B})$ is
  facet inducing for $\CutP(\K_{1,m_\A,m_\B})$ except for the cases
  that the inequality $\vct{a}^{\trans}\vct{x}\le0$ is a triangle
  inequality labelled as either
  $-x_{XA_1}-x_{XA_2}+x_{A_1A_2}\le0$ or
  $-x_{A_1A_2}-x_{A_1A_3}+x_{A_2A_3}\le0$.
\end{theorem}

For example, as we saw, the CHSH inequality is
the triangular elimination of Bell's original inequality, which is a triangle inequality.
The $I_{3322}$ inequality, found by Pitowsky and Svozil~\cite{PitSvo-PRA01} and Collins and Gisin~\cite{ColGis-JPA04},
is the triangular elimination of
a pentagonal inequality.

\begin{proof}

Let $r_{F}$ be the set of cut vectors on the hyperplane $\vct{a}^{\prime\trans} \vct{x}^{\prime} = 0$:
 $r_{F} =\left\{ \vct{\delta}^{\prime}(S^{\prime}) \mid \vct{a}^{\prime\trans} \vct{\delta}^{\prime} = 0, S^{\prime}:\text{cut} \right\} $
for $\CutP(\K_{1,m_\A,m_\B})$.
We prove the theorem by exhibiting a linearly independent subset of 
these cut vectors with cardinality $m_\A + m_\B + m_\A m_\B -1$.

In the following proof, we consider a simple case of $n_\B=1$.
We consider the other case later.
In addition, we assume that  $a_{A_{i}A_{i^{\prime}}} \leq 0$ for all
eliminated terms. For the other cases, the proof is similar.

By the above restriction, $m_\A + m_\B + m_\A m_\B -1 = (n_\A^3+3n_\A)/2$.

A sketch of proof is as follows: first, we restrict $r_F$ and decompose the whole space of
 $\CutP(\K_{1,m_\A,m_\B})$ into
two subspaces. For each subspace, we can pick a set of cut vectors
which are linearly independent in that subspace. Next, we show that these sets of cut vectors
are linearly independent in the whole space.

First, let the subset $r^{\prime}_{F}$ of $r_F$ be those cuts such that, for any $1 \leq i < i^{\prime} \leq n_\A$,
two vertices $A_{i^{\prime}}$ and $B'_{A_{i}A_{i^{\prime}}}$ are assigned same value.
Then, consider the intersection of the space spanned by $\vct{\delta}^{\prime}(S^{\prime}) \in r^{\prime}_{F}$ and the subspace
\[W = \left\{ ( x_{XA_{i}}, x_{XB_{j}},
x_{A_{i}B_{j}},
x_{A_{i}B'_{A_{i}A_{i^{\prime}}}})^\trans
  _{1 \leq i < i^{\prime} \leq n_\A, 1 \leq j \leq n_\B}
   \right\}. \]
From the definition of $r^{\prime}_{F}$, $\delta^{\prime}_{A_{i^{\prime}}B'_{A_{i}A_{i^{\prime}}}}(S^{\prime}) = 0$.
Therefore,
\begin{eqnarray*}
  \vct{a}^{\prime\trans} \vct{\delta}^{\prime}(S^{\prime}) 
  = \sum_{1 \leq i \leq n_\A} a_{XA_{i}} \delta_{XA_{i}} (S^{\prime}) + 
     \sum_{1 \leq j \leq n_\B} a_{XB_{j}} \delta_{XB_{j}} (S^{\prime}) \\
  + \sum_{1 \leq i \leq n_\A,1 \leq j \leq n_\B} a_{A_{i}B_{j}} \delta_{A_{i}B_{j}} (S^{\prime})
 + \sum_{1 \leq i < i^{\prime} \leq n_\A}  a_{A_{i}A_{i^{\prime}}} \delta^{\prime}_{A_{i}B'_{A_{i}A_{i^{\prime}}}} (S^{\prime}) = 0.
\end{eqnarray*}
This means that the intersection of space spanned by $\vct{\delta}^{\prime}(S^{\prime}) \in r^{\prime}_{F}$ and $W$ is equivalent to the space spanned by the cut vectors $r_{f} = \left\{ \vct{\delta}(S) \mid \vct{a}^\trans \vct{\delta} = 0, S:\text{cut} \right\} $
of $\CutP(\K_{1 + n_\A + n_\B})$. 
Therefore, from the assumption that the inequality $\vct{a}^{\trans} \vct{x} \leq 0$ is facet supporting,
we can pick   $(n_\A^2 + 3n_\A)/2$ linearly independent cut vectors and transform the cut vectors of $\CutP(\K_{1 + n_\A + n_\B})$ into
corresponding cut vectors of $\CutP(\K_{1,m_\A,m_\B})$. Let this set of linearly independent cut vectors be $D_{0}$.

The remaining subspace of $\CutP(\K_{1,m_\A,m_\B})$ is 
\[V = \bigoplus_{i < i^{\prime}} V_{A_{i}A_{i^{\prime}}} = \bigoplus_{i < i^{\prime} } \left\{ \left( x_{XB'_{A_{i}A_{i^{\prime}}}}, x_{A_{i^{\prime}}B'_{A_{i}A_{i^{\prime}}}},
 x_{A_{i^{\prime\prime}}B'_{A_{i}A_{i^{\prime}}}} \right)_{i^{\prime\prime} \neq i, i^{\prime}}^{\trans} \right\}  \]
for each eliminated term $A_{i}A_{i^{\prime}}, 1 \leq i < i^{\prime} \leq n_\A$.

Instead of $V$, we consider the space 
\[
  \fl
  V^{\prime} = \bigoplus_{i < i^{\prime}}  V^{\prime}_{A_{i}A_{i^{\prime}}}
  = \bigoplus_{i < i^{\prime}}
    \left\{\left(
      x_{XB'_{A_{i}A_{i^{\prime}}}}-x_{A_{i^{\prime}}B'_{A_{i}A_{i^{\prime}}}},
      x_{A_{i^{\prime}}B'_{A_{i}A_{i^{\prime}}}},
      x_{\alpha_{A_{i},A_{i^{\prime}},A_{i^{\prime\prime}} }}
    \right)_{i^{\prime\prime} \neq i, i^{\prime}}^{\trans} \right\}
\]
where
\[
  \fl
  x_{\alpha_{A_{i},A_{i^{\prime}},A_{i^{\prime\prime}}}}
  = \left\{ \begin{array}{ll}
     \frac{1}{2} ( x_{A_{i^{\prime\prime}} B'_{A_{i}A_{i^{\prime}} } }
                 - x_{A_{i^{\prime}} B'_{A_{i^{\prime}}A_{i^{\prime\prime}} } }
                 - x_{A_{i^{\prime}} B'_{A_{i}A_{i^{\prime}} } }
                 + 3x_{A_{i^{\prime\prime}} B'_{A_{i^{\prime}}A_{i^{\prime\prime}} } } )
                 & (i^{\prime} < i^{\prime\prime} ) \\
     \frac{1}{2}(x_{A_{i^{\prime\prime}} B'_{A_{i}A_{i^{\prime}} } }
                 - x_{A_{i^{\prime\prime}} B'_{A_{i^{\prime\prime}}A_{i^{\prime}} } }
                 - x_{A_{i^{\prime}} B'_{A_{i}A_{i^{\prime}} } }
                 - x_{A_{i^{\prime}} B'_{A_{i^{\prime\prime}}A_{i^{\prime}} } } )
                 & (i^{\prime\prime} < i^{\prime} )
  \end{array} \right.
\]
in the following.
Since the transform $V$ to $V^{\prime}$ is linear, the linear independence of vectors in $V$ 
is equivalent to that in $V^{\prime}$.

Then, we consider the subset $r^{\prime\prime}_{F,A_{i}A_{i^{\prime}} }$ of $r_F$ for each $A_{i}A_{i^{\prime}}$
restricted as follows:
$A_{i^{\prime}}$ must be assigned $0$ and both $B'_{A_{i}A_{i^{\prime}}}$ and $A_{i}$ must be assigned $1$.
For other terms $A_{i^{\prime\prime\prime}}A_{i^{\prime\prime\prime\prime}}(1 \leq i^{\prime\prime\prime} < i^{\prime\prime\prime\prime} \leq n_\A)$,
vertices $A_{i^{\prime\prime\prime\prime}}$ and $B'_{A_{i^{\prime\prime\prime}}A_{i^{\prime\prime\prime\prime}}}$ must be assigned the same value.
From that restriction, the equations
\begin{eqnarray*}
\delta^{\prime}_{XB'_{A_{i}A_{i^{\prime}}}}(S^{\prime\prime}) - \delta^{\prime}_{A_{i^{\prime}}B'_{A_{i}A_{i^{\prime}}}}(S^{\prime\prime}) = -\delta_{XA_{i^{\prime}}}(S), \\
\delta^{\prime}_{A_{i^{\prime}}B'_{A_{i}A_{i^{\prime}}}}(S^{\prime\prime}) = 1, \\
\delta^{\prime}_{\alpha_{A_{i}, A_{i^{\prime}}, A_{i^{\prime\prime}} }}(S^{\prime\prime}) = -\delta_{A_{i^{\prime\prime}}A_{i^{\prime}}}(S) 
\end{eqnarray*}
hold for $\vct{\delta}^{\prime}(S^{\prime\prime}) \in r^{\prime\prime}_{F,A_{i}A_{i^{\prime}}} $.
This means that the intersection of the space spanned by $\vct{\delta}^{\prime}(S^{\prime\prime})$ and
the subspace $V^{\prime}_{A_{i}A_{i^{\prime}}}$
is equivalent to that of the space spanned by $\vct{\delta}(S) \in r_{f}$ and the subspace
\[U_{A_{i}A_{i^{\prime}}}= \left\{ \left( x_{XA_{i^{\prime}}}, 1, x_{A_{i^{\prime\prime}}A_{i^{\prime}}} \right)_{i^{\prime\prime} \neq i, i^{\prime}}^{\trans} \right\}. \]

Now, because $r_f$ is on the hyperplane $\vct{a}^{\trans} \vct{x} = 0$, the above intersection has dimension  $n_\A$ or $n_\A-1$.
However, from the condition on the inequality $\vct{a}^\trans\vct{x}\le0$,
the space spanned
by $r_f$ is not parallel to $U_{A_{i}A_{i^{\prime}}}$. Therefore, the dimension is  $n_\A$ and
we can extract  $n_\A$ cut vectors which are linearly independent
in the subspace $V^{\prime}$ using the cut vectors from $r_{f}$. Let this set of cut vectors be
 $D_{A_{i}A_{i^{\prime}}}$.

Finally, we show that $D_0 \cup \bigcup_{1\le i<i'\le n_\A}D_{A_{i}A_{i^{\prime}}}$ is a linearly independent set of cut vectors.
Suppose that the linear combination
\[ \sum_{\vct{\delta}^{\prime\trans}(S^{\prime}) \in D_{0} } \kappa_{S^{\prime}}\vct{\delta}^{\prime\trans}(S^{\prime})  +
\sum_{1 \leq i < i^{\prime} \leq n_\A} \sum_{\vct{\delta}^{\prime\trans}(S^{\prime\prime}) \in D_{A_{i}A_{i^{\prime}}} } \lambda^{A_{i}A_{i^{\prime}}}_{S^{\prime\prime}}\vct{\delta}^{\prime\trans}(S^{\prime\prime})  = 0\]
holds.
Consider the subspace $V^{\prime}_{A_{i}A_{i^{\prime}}}$ of the above linear combination.
From the construction, for $D_0$ and $ D_{A_{i^{\prime\prime}}A_{i^{\prime\prime\prime}}} $,
the elements of cut vectors in that subspace are all zero.
Therefore, for the linear combination to hold, it must be that  $ \sum_{\vct{\delta}^{\prime\trans}(S^{\prime\prime}) \in D_{A_{i}A_{i^{\prime}}} } \lambda^{A_{i}A_{i^{\prime}}}_{S^{\prime\prime}}\vct{\delta}^{\prime\trans}(S^{\prime\prime})  = 0$.
However, the linear independence of $D_{A_{i}A_{i^{\prime}}}$ means that the coefficients are all zero.
By repeating this argument, we can conclude that the coefficient $ \lambda^{A_{i}A_{i^{\prime}}}_{S^{\prime\prime}}$ must be zero.
So, from the linear independence of $D_0$, the coefficients $ \kappa_{S^{\prime}}$ are also zero. 
This completes the proof for the case $n_\B=1$.

Now we describe the outline of the proof for general case.
The idea of the proof is to perform triangular elimination in two
steps: eliminate the edges $\A_i\A_{i'}$ for $1\le i<i'\le n_\A$ in
one step and then the edges $\B_j\B_{j'}$ for $1\le j<j'\le n_\B$ in
the other.
To do this, we need the notion of the cut polytope
$\CutP(G)\subseteq\RR^E$ of a general graph $G=(V,E)$, which is
obtained from the cut polytope of the complete graph on node set $V$
by removing the coordinates corresponding to the edges missing in $E$.
In particular we consider the cut polytopes of the following two
intermediate graphs: the graph $G_1(n_\A,n_\B)$ obtained from
$\K_{1,n_\A,m_\B}$ by adding edges $\B_j\B_{j'}$ and
$\B_j\B'_{\A_i\A_{i'}}$, and the graph $G_2(n_\A,n_\B)$ obtained from
$\K_{1,m_\A,m_\B}$ by adding edges $\B_j\B'_{\A_i\A_{i'}}$.

The next lemma is a basic fact from polytope theory (see
Lemma~26.5.2~(ii) in~\cite{DezLau:cut97};
though the statement there restricts $G$ to be a complete graph,
that restriction is not necessary).

\begin{lemma} \label{lemm:zero-proj}
  Let $G$ be a graph and $G'$ be a subgraph of $G$.
  If $\vct{a}^\trans\vct{x}\le0$ is facet inducing for $\CutP(G)$ and
  $a_e=0$ for all edges $e$ belonging to $G$ but not to $G'$, then
  $\vct{a}^\trans\vct{x}\le0$ is facet inducing also for $\CutP(G')$.
\end{lemma}

The inequality after the first step of triangular elimination is as
follows:
\begin{eqnarray}
  \fl
    \sum_{ 1 \leq i \leq n_\A}a_{XA_{i}}x_{XA_{i}}
  + \sum_{ 1 \leq j \leq n_\B}a_{XB_{j}}x_{XB_{j}}
  + \sum_{ 1 \leq i \leq n_\A,  1 \leq j \leq n_\B}
      a_{A_{i}B_{j}}x_{A_{i}B_{j}} \nonumber\\
  \fl
  + \sum_{ 1 \leq i < i^{\prime} \leq n_\A}
      ( a_{A_{i}A_{i^{\prime}}}x_{A_{i}B'_{A_{i}A_{i^{\prime}}}}
       - |a_{A_{i}A_{i^{\prime}}}| x_{A_{i^{\prime}}B'_{A_{i}A_{i^{\prime}}}}
      )
  +\sum_{ 1 \leq j < j^{\prime} \leq n_\B}
      a_{B_{j}B_{j^{\prime}}}x_{B_{j}B_{j^{\prime}}}
   \leq a_0. \label{eq:elimination-alice}
\end{eqnarray}

For the case $n_\B=1$, the inequality (\ref{eq:elimination-alice}) is
exactly the same as (\ref{eq:after-elimination}).
We proved above for the case $n_\B=1$ that the inequality
(\ref{eq:elimination-alice}) is facet inducing for
$\CutP(\K_{1,n_\A,m_\B})$.
Except for when the original inequality is a triangle inequality, we
can extend this argument to prove that the inequality
(\ref{eq:elimination-alice}) is facet inducing also for
$\CutP(G_1(n_\A,n_\B))$.
This can be generalized for the case $n_\B>1$: the inequality
(\ref{eq:elimination-alice}) is facet inducing for
$\CutP(G_1(n_\A,n_\B))$.
Then we can repeat a similar argument to prove the final inequality
(\ref{eq:after-elimination}) is facet inducing for
$\CutP(G_2(n_\A,n_\B))$.
Since $G_2(n_\A,n_\B)$ is a supergraph of the desired graph
$\K_{1,m_\A,m_\B}$, the inequality (\ref{eq:after-elimination}) is
facet inducing also for $\CutP(\K_{1,m_\A,m_\B})$ from
Lemma~\ref{lemm:zero-proj}.
\end{proof}

   \subsection{Triangular elimination and symmetry}\label{symmetry-of-cut-polytope}

Many Bell inequalities are equivalent to each other
due to the arbitrariness
in the labelling of the party, observable and value identifiers.
This corresponds to symmetries of the underlying polytope.
We consider ways of representing nonequivalent Bell inequalities in this section.

The nonequivalence of Bell inequalities can
be translated into two questions about facet inequalities $f$ and $f^{\prime}$
of a given
cut polytope of a complete graph, and their triangular eliminations  $F$ and $F^{\prime}$, respectively:
\begin{enumerate}
   \item does the equivalence of $f$ and $f^{\prime}$ imply the equivalence of $F$ and $F^{\prime}$?
   \item does the equivalence of $F$ and $F^{\prime}$ imply the equivalence of $f$ and $f^{\prime}$?
\end{enumerate}
The answers are both affirmative if we define equivalence appropriately,
so equivalence before triangular elimination is logically equivalent
to equivalence after triangular elimination.
This means that, for example, to enumerate the nonequivalent Bell inequalities, we need only enumerate
the facet inequalities of the cut polytope of the complete graph 
up to symmetry by party, observable and value exchange.

In $\CutP(\K_{1,m_\A,m_\B})$, the relabelling of all vertices of Alice to that of Bob  
and vice versa corresponds to a party exchange.
On the other hand, the local relabelling of some vertices of Alice (or Bob) corresponds to an observable exchange.
Thus by the observable exchange of Alice represented by the permutation $\sigma$ over $\{A_{1}, \ldots, A_{m_\A}\}$,
an inequality $\vct{a}^{\trans} \vct{x} \leq a_{0}$ is transformed into $\vct{a}^{\prime\trans} \vct{x} \leq a_{0}$ 
where $a^{\prime}_{\sigma(A_{i})V} = a_{A_{i}V}$ for any vertex $V$. 

In addition, there is an operation which corresponds to a value exchange of some observables, 
called a \emph{switching} in the theory of cut polytopes.
By the switching corresponding to the value exchange  of an Alice's observable $A_{i_{0}}$,
an inequality $\vct{a}^{\trans} \vct{x} \leq a_{0}$ is transformed into
$\vct{a}^{\prime\trans} \vct{x} \leq a_{0} - \sum_{V}a_{A_{i_{0}}V }$ where 
$a^{\prime}_{A_{i_{0}}V} = -a_{A_{i_{0}}V}$, and $a^{\prime}_{A_{i}V} =a_{A_{i}V}$
for any $i \neq i_{0}$ and any vertex $V \neq A_{i_{0}}$ (definitions for Bob's exchange are similar).

It is well known, and easily shown, that by repeated application of the  switching operation
we may reduce the right hand side of any facet inequality to zero.

Let $n_\A\le n_\B$ and $n=1+n_\A+n_\B$.
Let $f$ and $f'$ be facets of $\CutP_n$ where the $n$ nodes of $\K_n$
is labelled by $V=\{A_1,\dots,A_{n_\A},\allowbreak
B_1,\dots,B_{n_\B},\allowbreak X\}$.
The two facets $f$ and $f'$ are said to be \emph{equivalent} and
denoted $f\sim f'$ if $f$ can be transformed to $f'$ by applying zero
or more of the following operations:
(1) (only applicable in the case $n_\A=n_\B$) swapping labels of nodes
    $A_i$ and $B_i$ for all $1\le i\le n_\A$,
(2) relabelling the nodes within $A_1,\dots,A_{n_\A}$,
(3) relabelling the nodes within $B_1,\dots,B_{n_\B}$, and
(4) switching.%
\footnote{
The two facets $f$ and $f'$ are said to be \emph{equivalent} and
denoted $f\sim f'$ if $f$ can be transformed to $f'$ by permutation
and switching where the permutation $\tau$ on $V$ satisfies:
(1) $\tau(X)=X$ and (2) $\tau$ either fixes two sets
$\{A_1,\dots,A_{n_\A}\}$ and $\{B_1,\dots,B_{n_\B}\}$ setwise or (in
the case $n_\A=n_\B$) swaps these two sets.
}

Two facets $F$ and $F'$ of $\CutP(\K_{1,m_\A,m_\B})$ are said to be
\emph{equivalent} and denoted $F\sim F'$ if $F$ can be transformed to
$F'$ by applying permutation which fixes node $X$, switching, or
both.
This notion of equivalence of facets of $\CutP(\K_{1,m_\A,m_\B})$
corresponds to equivalence of tight Bell inequalities up to party,
observable and value exchange.

\begin{theorem} \label{thrm:sw-perm-n-even}
Let the triangular elimination of facet inequalities $f$ and $f^{\prime}$ be $F$ and $F^{\prime}$, respectively. Then,
$f \sim f^{\prime}$ $\iff$ $F \sim F^{\prime}$.
\end{theorem}
\begin{proof}
A sketch of the proof is as follows.
Since the permutation and switching operations are commutative,
it is sufficient to prove the
proposition under each operation separately.
Because the $\Rightarrow$ direction is straightforward for both permutation and switching,
we concentrate on the proof of $\Leftarrow$ direction.

First, consider switching. Suppose $F$ is obtained from a switching of
$F^{\prime}$. The switching could involve either (i) a new observable introduced
by the triangular elimination, or (ii) an observable which had a joint measurement term eliminated.
Since a switching of type (i) has no effect on $f$ and $f^{\prime}$,
we need only consider type (ii). We can view the triangular elimination of the term $A_{i}A_{i^{\prime}}$
as addition of triangle inequality $x_{A_{i}A_{i^{\prime}}} - x_{A_{i}B'_{A_{i}A_{i^{\prime}}}} - x_{A_{i^{\trans}}B'_{A_{i}A_{i^{\prime}}}} \leq 0$
or its switching equivalent inequality $- x_{A_{i}A_{i^{\prime}}} - x_{A_{i}B'_{A_{i}A_{i^{\prime}}}} + x_{A_{i^{\trans}}B'_{A_{i}A_{i^{\prime}}}} \leq 0$ according to the sign of the coefficient $a_{A_{i}A_{i^{\prime}}}$.
Thus, if $F$ is switching of $F^{\prime}$ of vertices $A_{i}$ and $B'_{A_{i}A_{i^{\prime}}}$ then $f$ is switching of $f^{\prime}$ of $A_{i}$.

Next, consider the permutation corresponding to an observable exchange.
Observe that for any vertex $A_{i} (1 \leq i \leq n_\A)$, triangular elimination
does not change the number of terms $A_{i}V$ with non-zero coefficient. In addition,
it can be shown that for any facet inequality $f$ of the cut polytope of the complete graph other than the triangle inequality,
there is no vertex satisfying the following conditions: (a) there are exactly two terms $A_{i}V$,
 with non-zero coefficients, and (b)
for those non-zero coefficients $a_{A_{i}W}$ and $a_{A_{i}U}$, $|a_{A_{i}W}| = |a_{A_{i}U}|$~\cite{AviImaItoSas:0404014}.
This means that if $F \sim F^{\prime}$, then the corresponding permutation $\sigma$ is always in the following form:
for permutations $\tau_{A}$ over $\{ A_{1}, \ldots, A_{n_\A}\}$ and $\tau_{B}$ over $\{ B_{1}, \ldots, B_{n_\B}\}$,
$\sigma(A_{i}) = \tau_{A}(A_{i})$ and $\sigma(B'_{A_{i}A_{i^{\prime}}}) = B_{\tau_{A}(A_{i})\tau_{A}(A_{i^{\prime}})}$.
The situation is the same for Bob. 

Therefore, $f$ and $f^{\prime}$ are equivalent under the permutations $\tau_{A}$ and $\tau_{B}$.
\end{proof}

\subsection{Computational results}

\begin{table}
  \caption{\label{table:count}
    The number of inequivalent facets of $\CutP_n$ and
    the number of inequivalent tight Bell inequalities
    obtained as the triangular eliminations of the facets of
    $\CutP_n$.
    Asterisk (*) indicates the value is a lower bound.}
  \begin{indented} \item
  \begin{tabular}{@{}ccc} \br
    $n$ & Facets of $\CutP_n$ &
                     Tight Bell ineqs.\ via triangular elimination \\ \mr
     3  &       1  &           2  \\
     4  &       1  &           2  \\
     5  &       2  &           8  \\
     6  &       3  &          22  \\
     7  &      11  &         323  \\
     8  &     147* &      40,399* \\
     9  & 164,506* & 201,374,783* \\ \br
  \end{tabular}
  \end{indented}
\end{table}

By Theorem~\ref{thrm:sw-perm-n-even},
we can compute the number of the classes of facets of
$\CutP(\K_{1,m_A,m_B})$
of the same type obtained by applying triangular elimination
to non-triangular facets of $\CutP_n$.
We consulted De~Simone, Deza and Laurent~\cite{DesDezLau-DM94}
for the H-representation of $\CutC_7$,
and the ``conjectured complete description'' of $\CutC_8$
and the ``description possibly complete'' of $\CutC_9$ in SMAPO~\cite{SMAPO}.
The result is summarized in \tref{table:count}.
For $n=8$ and $9$, the number is a lower bound since the known
list of the facets of $\CutP_n$ is not proved to be complete.
A program to generate Bell inequalities from the list in \cite{SMAPO}
are available from an author's webpage at
\url{http://www-imai.is.s.u-tokyo.ac.jp/~tsuyoshi/bell/}.
The list of the generated Bell inequalities for $n=8$ is also
available.

\section{Families of Bell inequalities} \label{sect:families}

While a large list of individual tight Bell inequalities is useful in
some applications, a few formulas which give many different Bell
inequalities for different values of parameters are easier to treat
theoretically.
The cut polytope of the complete graph has several classes of valid
inequalities whose subclasses of facet-inducing inequalities are
partially known (see \cite[Chapters~27--30]{DezLau:cut97} for
details).
In this section, we apply triangular elimination to two typical
examples of such classes to obtain two general formulas for Bell
inequalities.
In addition, we prove sufficient conditions for these formulas to give
a tight Bell inequality.

In this section, terms of the left
hand side of an inequality are arrayed in the format introduced by
Collins and Gisin~\cite{ColGis-JPA04}; each row corresponds to
coefficients of each observable of party $A$ and each column
corresponds to that of party $B$.  Because of switching equivalence,
we can assume that the right hand side of inequality are always zero.
The example of the CHSH $ -q_{A_{1}} -q_{B_{1}} +
q_{A_{1}B_{1}} + q_{A_{1}B_{2}} + q_{A_{2}B_{1}} - q_{A_{2}B_{2}} \leq 0$
is arrayed in the form as follows:
\[
  \left( \begin{array}{c||cc}
    &-1 & 0\\ \hline \\[-14pt] \hline
  -1& 1 & 1\\
   0& 1 &-1
  \end{array} \right) \le0.
\]

\subsection{Bell inequalities derived from hypermetric inequalities}

\emph{Hypermetric inequalities} are a fundamental class of
inequalities valid for the cut polytope of the complete graph.
Here we derive a new family of Bell inequalities by applying
triangular elimination to the hypermetric inequalities.
 A special case of this
 family, namely the triangular eliminated pure hypermetric inequality,
 contains four previously known Bell inequalities: the trivial
 inequalities like $q_{A_{1}} \leq 1$, the well
known CHSH inequality found by Clauser, Horne, Shimony and
Holt~\cite{ClaHorShiHol-PRL69}, the inequality named $I_{3322}$ by
Collins and Gisin~\cite{ColGis-JPA04},
originally found by Pitowsky
and Svozil~\cite{PitSvo-PRA01}, and the $I_{3422}^2$ inequality by
Collins and Gisin~\cite{ColGis-JPA04}.

Let $s$ and $t$ be nonnegative integers and
$b_{\A_1},\dots,b_{\A_s},\allowbreak
 b_{\B_1},\dots,b_{\B_t}$ be integers.
We define $b_{\X}=1-\sum_{i=1}^s b_{\A_i}-\sum_{j=1}^t b_{\B_j}$.
Then it is known that $\sum_{uv} b_u b_v x_{uv}\le0$, where the sum is
taken over the $\binom{s+t+1}{2}$ edges of the complete graph on nodes
$\X,\A_1,\dots,\A_s,\allowbreak \B_1,\dots,\B_t$, is valid for
$\CutP_{s+t+1}$.
This inequality is called the \emph{hypermetric inequality} defined by
the weight vector $\vct{b}=(b_{\X},\allowbreak
b_{\A_1},\dots,b_{\A_s},\allowbreak
b_{\B_1},\dots,b_{\B_t})$.

We apply triangular elimination to this hypermetric inequality.
Let $s_+$ and $t_+$ be the number of positive entries of the form
$b_{\A_i}$ and of the form $b_{\B_j}$, respectively.
Without loss of generality, we assume that
$b_{\A_1},\dots,b_{\A_{s_+}},\allowbreak
 b_{\B_1},\dots,b_{\B_{t_+}}>0$, and
$b_{\A_{s_++1}},\dots,b_{\A_s},\allowbreak
 b_{\B_{t_++1}},\dots,b_{\B_t}\le0$.
By assigning $a_{uv}=b_ub_v$ in the
formula~(\ref{eq:after-elimination}), the Bell inequality obtained by
triangular elimination is:
\begin{eqnarray}
   \sum_{i=1}^{s_+}b_{\A_i}\biggl(\frac{1-b_{\A_i}}{2}-\sum_{i'=1}^{i-1}b_{\A_{i'}}\biggr)q_{\A_i}
  +\sum_{i=s_++1}^s b_{\A_i}\biggl(\frac{1-b_{\A_i}}{2}-\sum_{i'=s_++1}^{i-1}b_{\A_{i'}}\biggr)q_{\A_i} \nonumber\\
  +\sum_{j=1}^{t_+}b_{\B_j}\biggl(\frac{1-b_{\B_j}}{2}-\sum_{j'=1}^{j-1}b_{\B_{j'}}\biggr)q_{\B_j}
  +\sum_{j=t_++1}^t b_{\B_j}\biggl(\frac{1-b_{\B_j}}{2}-\sum_{j'=t_++1}^{j-1}b_{\B_{j'}}\biggr)q_{\B_j} \nonumber\\
  +\sum_{j=1}^{t_+}\sum_{j'=t_++1}^t b_{\B_j}b_{\B_{j'}}q_{\A'_{jj'}}
  +\sum_{i=1}^{s_+}\sum_{i'=s_++1}^s b_{\A_i}b_{\A_{i'}}q_{\B'_{ii'}}
  -\sum_{i=1}^s\sum_{j=1}^t b_{\A_i}b_{\B_j}q_{\A_i\B_j} \nonumber\\
  -\sum_{1\le i<i'\le s}b_{\A_i}b_{\A_{i'}}q_{\A_i\B'_{ii'}}
  +\sum_{1\le i<i'\le s}\lvert b_{\A_i}b_{\A_{i'}}\rvert q_{\A_{i'}\B'_{ii'}}
  \nonumber\\
  -\sum_{1\le j<j'\le t}b_{\B_j}b_{\B_{j'}}q_{\A'_{jj'}\B_j}
  +\sum_{1\le j<j'\le t}\lvert b_{\B_j}b_{\B_{j'}}\rvert q_{\A'_{jj'}\B_{j'}}
  \le0.
  \label{eq:hypermetric-bell}
\end{eqnarray}

Though the formula~(\ref{eq:hypermetric-bell}) represents a Bell
inequality for any choice of weight vector $\vct{b}$, this Bell
inequality is not always tight.
Many sufficient conditions for a hypermetric inequality to be
facet-inducing are known in study of cut polytopes.
By Theorem~\ref{thrm:kn-k1mm}, these sufficient conditions give
sufficient conditions for the Bell
inequality~(\ref{eq:hypermetric-bell}) to be tight.
The sufficient conditions stated in
\cite[Corollary~27.2.5]{DezLau:cut97} give the following theorem.

\begin{theorem} \label{thrm:hypermetric-bell-facet}
  The Bell inequality~(\ref{eq:hypermetric-bell}) is tight if one of
  the following conditions is satisfied.
  \begin{enumerate}[(i)]
    \item \label{enum:pure-hypermetric}
      For some $l>1$, the integers
      $b_{\A_1},\dots,b_{\A_s},\allowbreak b_{\B_1},\dots,b_{\B_t}$
      and $b_{\X}$ contain $l+1$ entries equal to $1$ and $l$ entries
      equal to $-1$, and the other entries (if any) are equal to $0$.
    \item
      At least $3$ and at most $n-3$ entries in
      $b_{\A_1},\dots,b_{\A_s},\allowbreak b_{\B_1},\dots,b_{\B_t}$
      and $b_{\X}$ are positive, and all the other entries are equal
      to $-1$.
  \end{enumerate}
\end{theorem}

Now we consider some concrete cases when the
formula~(\ref{eq:hypermetric-bell}) represents a tight Bell
inequality.
If we let $s+t=2l$, $s\le l$, $l>1$,
$b_{\A_1}=\dots=b_{\A_s}=b_{\B_1}=\dots=b_{\B_{l-s}}=1$, and
$b_{\B_{l-s+1}}=\dots=b_{\B_t}=-1$, then $b_{\X}=1$ and by
case~(\ref{enum:pure-hypermetric}) of
Theorem~\ref{thrm:hypermetric-bell-facet}, the Bell
inequality~(\ref{eq:hypermetric-bell}) is tight.
In this case, the Bell inequality~(\ref{eq:hypermetric-bell}) is in
the following form.
\begin{eqnarray}
  \fl
  -\sum_{i=1}^s(i-1)q_{\A_i}
  -\sum_{j=1}^{l-s}\sum_{j'=l-s+1}^t q_{\A'_{jj'}}
  -\sum_{j=1}^{l-s}(j-1)q_{\B_j}
  -\sum_{j=l-s+1}^t(j-(l-s))q_{\B_j} \nonumber\\
  \fl
  -\sum_{i=1}^s\sum_{j=1}^{l-s} q_{\A_i\B_j}
  +\sum_{i=1}^s\sum_{j=l-s+1}^t q_{\A_i\B_j}
  -\sum_{1\le i<i'\le s}q_{\A_i\B'_{ii'}}
  +\sum_{1\le i<i'\le s}q_{\A_{i'}\B'_{ii'}} \nonumber\\
  \fl
  -\sum_{1\le j<j'\le l-s}q_{\A'_{jj'}\B_j}
  -\sum_{l-s+1\le j<j'\le t}q_{\A'_{jj'}\B_j}
  +\sum_{j=1}^{l-s}\sum_{j'=l-s+1}^t q_{\A'_{jj'}\B_j}
  +\sum_{1\le j<j'\le t}q_{\A'_{jj'}\B_{j'}} \le0.
  \label{eq:pure-hypermetric-bell-1}
\end{eqnarray}

Examples of tight Bell inequality in the
form~(\ref{eq:pure-hypermetric-bell-1}) are $\I_{3322}$ and
$\I_{3422}^2$ inequalities~\cite{ColGis-JPA04}.

In case of $l=1$, Theorem~\ref{thrm:hypermetric-bell-facet} does not
guarantee that the Bell inequality~(\ref{eq:pure-hypermetric-bell-1})
is tight.
However, in cases of $(l,s,t)=(1,1,1)$ and $(1,1,2)$, the Bell
inequality~(\ref{eq:pure-hypermetric-bell-1}) becomes trivial and CHSH
inequalities, respectively, both of which are tight.

Letting $(l,s,t)=(2,2,2)$ in (\ref{eq:pure-hypermetric-bell-1}) gives:
\begin{equation}
  \fl
  -q_{\A_2}-q_{\B_1}-2q_{\B_2}
  +q_{\A_1\B_1}+q_{\A_1\B_2}+q_{\A_2\B_1}+q_{\A_2\B_2}
  -q_{\A_1\B'_{12}}+q_{\A_2\B'_{12}}-q_{\A'_{12}\B_1}+q_{\A'_{12}\B_2} \le0.
  \label{eq:i3322-1}
\end{equation}
Following the notation in~\cite{ColGis-JPA04}, we write the
inequality~(\ref{eq:i3322-1}) by arraying its coefficients:
\[
  \left(\begin{array}{cc||ccc}
               &    & (\A_2) & (\A_1) & (\A'_{12}) \\
               &    &   -1   &    0   &     0      \\ \hline\\[-14pt]\hline
    (\B_2)     & -2 &    1   &    1   &     1      \\
    (\B_1)     & -1 &    1   &    1   &    -1      \\
    (\B'_{12}) &  0 &    1   &   -1   &     0
  \end{array}\right) \le 0.
\]
Now it is clear that the Bell inequality~(\ref{eq:i3322-1}) is
$\I_{3322}$ inequality.

Letting $(l,s,t)=(2,1,3)$ in (\ref{eq:pure-hypermetric-bell-1})
gives:
\begin{equation}
  \left(\begin{array}{cc||ccc}
               &    & (\B_2) & (\B_3) & (\B_1) \\
               &    &   -1   &   -2   &    0   \\ \hline\\[-14pt]\hline
    (\A_1)     &  0 &    1   &    1   &   -1   \\
    (\A'_{13}) & -1 &    0   &    1   &    1   \\
    (\A'_{12}) & -1 &    1   &    0   &    1   \\
    (\A'_{23}) &  0 &   -1   &    1   &    0
  \end{array}\right) \le 0.
  \label{eq:i34222-1}
\end{equation}
After exchanging the two values $1$ and $0$ of the observable
$\A_1$, and doing the same to the two values of the observable $\B_3$,
the Bell inequality~(\ref{eq:i34222-1}) becomes:
\[
  \left(\begin{array}{cc||ccc}
               &    & (\B_2) & (\overline{\B_3})
                                      & (\B_1) \\
               &    &    0   &    1   &   -1   \\ \hline\\[-14pt]\hline
    (\overline{\A_1})
               & -1 &   -1   &    1   &    1   \\
    (\A'_{13}) &  0 &    0   &   -1   &    1   \\
    (\A'_{12}) & -1 &    1   &    0   &    1   \\
    (\A'_{23}) &  1 &   -1   &   -1   &    0
  \end{array}\right) \le 1,
\]
which is $\I_{3422}^2$ inequality~\cite{ColGis-JPA04}.
This means that the Bell inequality~(\ref{eq:i34222-1}) is equivalent
to $\I_{3422}^2$ inequality.

\subsection{Bell inequalities derived from pure clique-web inequalities}

Clique-web inequalities~\cite[Chapter~29]{DezLau:cut97} are
generalization of hypermetric inequalities.
One of the important subclasses of clique-web inequalities are the pure
clique-web inequalities, which are always facet-inducing.
Here we introduce an example of Bell inequalities
derived from some pure clique-web inequalities.

For nonnegative integers $s$, $t$ and $r$ with $s\ge t\ge2$ and
$s-t=2r$, we consider the pure clique-web inequality with parameters
$n=s+t+1$, $p=s+1$, $q=t$ and $r$.
After relabelling the $n$ vertices of $\K_n$ by
$\A_1,\dots,\A_s,\X,\B_1,\dots,\B_t$ in this order, the Bell
inequality~(\ref{eq:after-elimination}) corresponding to the clique-web
inequality is:
\begin{eqnarray}
  -\sum_{i=r+1}^{s-r}(i-r-1)q_{\A_i}-2t\sum_{i=s-r+1}^s q_{\A_i}
  -\sum_{j=1}^t (j-r)q_{\B_j}
  +\sum_{i=1}^s\sum_{j=1}^t q_{\A_i\B_j} \nonumber\\
  +\sum_{\substack{1\le i<i'\le s \cr r+1\le j-i\le s-r}} 
    (-q_{\A_i\B'_{ii'}}+q_{\A_{i'}\B'_{ii'}})
  +\sum_{1\le j<j'\le t}(-q_{\A'_{jj'}\B_j}+q_{\A'_{jj'}\B_{j'}})\le0.
  \label{eq:pure-cw-bell-1}
\end{eqnarray}

The next theorem is a direct consequence of
Theorem~\ref{thrm:kn-k1mm}.

\begin{theorem} \label{thrm:pure-cw-bell-facet}
  For any nonnegative integers $s$, $t$ and $r$ with $s\ge t\ge2$ and
  $s-t=2r$, the Bell inequality~(\ref{eq:pure-cw-bell-1}) is tight.
\end{theorem}

\subsection{Inclusion relation}

Collins and Gisin~\cite{ColGis-JPA04} pointed out that the following
$\I_{3322}$ inequality becomes the CHSH inequality if we fix two
measurements $\A_3$ and $\B_1$ to a deterministic measurement whose
result is always $0$.
\begin{eqnarray*}
  \text{$\I_{3322}$:} \quad &
  \left(\begin{array}{cc||ccc}
              &        & (\A_1) & (\A_2) & (\A_3) \\
              &        &   -1   &    0   &    0   \\ \hline\\[-14pt]\hline
    (\B_1)    &   -2   &    1   &    1   &    1   \\
    (\B_2)    &   -1   &    1   &    1   &   -1   \\
    (\B_3)    &    0   &    1   &   -1   &    0
  \end{array} \right)\le0, \\
  \text{CHSH:} \quad &
  \left(\begin{array}{cc||cc}
              &        & (\A_1) & (\A_2) \\
              &        &   -1   &    0   \\ \hline\\[-14pt]\hline
    (\B_2)    &   -1   &    1   &    1   \\
    (\B_3)    &    0   &    1   &   -1
  \end{array} \right)\le0.
\end{eqnarray*}
As stated in \cite{ColGis-JPA04}, this fact implies the CHSH
inequality is irrelevant if the $\I_{3322}$ inequality is given.
In other words, if a quantum state satisfies the $\I_{3322}$
inequality with every set of  measurements, then it also satisfies the CHSH
inequality with every set of measurements.

We generalize this argument and define \emph{inclusion relation}
between two Bell inequalities:
A Bell inequality $\vct{a}^\trans\vct{q}\le0$ \emph{includes}
another Bell inequality $\vct{b}^\trans\vct{q}\le0$ if we can obtain
the inequality $\vct{b}^\trans\vct{q}\le0$ by fixing some
measurements in the inequality $\vct{a}^\trans\vct{q}\le0$ to
deterministic ones.

We do not know whether all the Bell inequalities (except positive probability) include the CHSH
inequality.
However, we can prove that many Bell inequalities represented by
(\ref{eq:hypermetric-bell}) or (\ref{eq:pure-cw-bell-1}) include the
CHSH inequality.

\begin{theorem} \label{thrm:hypermetric-bell-facet-chsh}
  If $b_{\A_1}=b_{\A_2}=1$ and $b_{\B_{t_++1}}=-1$, then the Bell
  inequality represented by~(\ref{eq:hypermetric-bell}) contains the
  CHSH inequality.
\end{theorem}

\begin{proof}
  The Bell inequality~(\ref{eq:hypermetric-bell}) contains
  $s+\binom{t}{2}$ observables of Alice and $t+\binom{s}{2}$
  observables of Bob.
  By fixing all but 4 observables $\A_1$, $\A_2$, $\B_{t_++1}$ and
  $\B'_{12}$ to the one whose value is always $0$, we obtain the
  following CHSH inequality:
  $
    -q_{\A_2}-q_{\B_{t_++1}}
    +q_{\A_1\B_{t_++1}}+q_{\A_2\B_{t_++1}}
    -q_{\A_1\B'_{12}}+q_{\A_2\B'_{12}}\le0
  $.
\end{proof}

\begin{theorem} \label{thrm:pure-cw-bell-facet-chsh}
  All the Bell inequalities in the form~(\ref{eq:pure-cw-bell-1})
  include the CHSH inequality.
\end{theorem}

\begin{proof}
  By fixing all but 4 observables $\A_{r+1}$, $\A_{r+2}$, $\B_{r+1}$
  and $\B'_{r{+}1,r{+}2}$ to the one whose value is always $0$, the
  Bell inequality~(\ref{eq:pure-cw-bell-1}) becomes the following CHSH
  inequality:
  $
    -q_{\A_{r+2}}-q_{\B_{r+1}}
    +q_{\A_{r+1}\B_{r+1}}+q_{\A_{r+2}\B_{r+1}}
    -q_{\A_{r+1}\B'_{r{+}1,r{+}2}}+q_{\A_{r+2}\B'_{r{+}1,r{+}2}}\le0
  $.
\end{proof}


\subsection{Relationship between $I_{mm22}$ and triangular eliminated Bell inequality}
Collins and Gisin~\cite{ColGis-JPA04} proposed a family of tight Bell
inequalities obtained by the extension of CHSH and $I_{3322}$ as
$I_{mm22}$ family, and conjectured that $I_{mm22}$ is always facet
supporting (they also confirmed that for $m \leq 7$, $I_{mm22}$ is
actually facet supporting by computation). Therefore, whether their
$I_{mm22}$ can be obtained by triangular elimination of some facet
class of $\CutP(\K_{n})$ is an interesting question.

The $I_{mm22}$ family has the structure as follows:
\[
        \left( \begin{array}{c||cccccc}
      &  -1  &   0  &  \cdots &    0    &    0    & 0\\ \hline\\[-14pt]\hline
-(m-1)&   1  &   1  &  \cdots &    1    &    1    & 1\\
-(m-2)&   1  &   1  &  \cdots &    1    &    1    &-1\\
-(m-3)&   1  &   1  &  \cdots &    1    &   -1    & 0\\
\vdots&\vdots&\vdots&\revddots&\revddots&\revddots&\vdots\\
   -1 &   1  &   1  &   -1    &    0    &  \cdots & 0\\
    0 &   1  &  -1  &    0    &    0    &  \cdots & 0
        \end{array} \right)\le0.
\]

From its structure, it is straightforward that if $I_{mm22}$ can be obtained by
triangular elimination of some facet class of $\CutP_{n}$, then only
$A_{m}$ and $B_{m}$ are new vertices introduced by triangular elimination,
since the other vertices have degree more than $2$.
For $m=2,3,4$, the $I_{mm22}$ inequality is the triangular elimination of
the triangle, pentagon and Grishukhin
inequality $\sum_{1 \leq i < j \leq 4}x_{ij} +x_{56} +x_{57} -x_{67} -x_{16} -x_{36}
-x_{27} -x_{47} - 2 \sum_{1 \leq i \leq 4}x_{i5} \leq 0$, respectively.
In general, $I_{mm22}$ inequality is the triangular elimination of a
facet-inducing inequality of $\CutP_{2m-1}$ and it is
tight~\cite{AviIto-JH05}.

\subsection{Known tight Bell inequalities other than the triangular
  elimination of $\CutP(\K_{n})$}

Since we have obtained a large number of tight Bell inequalities by
triangular elimination of $\CutP(\K_{n})$, the next question is
whether they are complete i.e., whether all families and their
equivalents form the whole set of facets of $\CutP(\K_{1,m_\A,m_\B})$.

For the case $m_\A=m_\B=3$, the answer is affirmative.
Both \'{S}liwa~\cite{Sli-PLA03}
and Collins and Gisin~\cite{ColGis-JPA04} showed that there are only
three kinds of inequivalent facets: positive probabilities, CHSH and $I_{3322}$,
corresponding to the triangle facet, the triangular elimination of the
triangle facet and the triangular elimination of the pentagonal facet
of $\CutP(\K_{n})$, respectively.

On the other hand, in the case $m_\A=3$ and $m_\B=4$, the answer is
negative.
Collins and Gisin enumerated all of the tight
Bell inequalities
and classified them into 6 families of equivalent inequalities~\cite{ColGis-JPA04}.
While positive probabilities, CHSH, $I_{3322}$ and $I_{3422}^2$ inequalities are
either facets of $\CutP(\K_{n})$ or their triangular eliminations, the other
two are not:
\[
\fl
I^{1}_{3422} =
        \left( \begin{array}{c||ccc}
  & 1& 1&-2\\ \hline\\[-14pt]\hline
 1&-1&-1& 1\\
 0&-1& 1& 1\\
 0& 1&-1& 1\\
 1&-1&-1&-1
        \end{array}\right)\le2, \qquad
I^{3}_{3422} =
        \left( \begin{array}{c||ccc}
  & 1& 0&-1\\ \hline\\[-14pt]\hline
 0&-2& 1& 1\\
 0& 0&-1& 1\\
-1& 1& 1& 1\\
 2&-1&-1&-1 
        \end{array}\right)\le2.
\]

\section{Concluding remarks} \label{sect:concluding}

We introduced triangular elimination to derive tight Bell inequalities
from the facet inequalities of the cut polytope of the complete graph.
Though it does not give the complete list of Bell inequalities, this
method derives not only many individual tight Bell inequalities from
individual known facet inequalities of cut polytope, but also several
families of Bell inequalities.
This gives a partial answer to the $N=K=2$ case of the problem posed
by Werner~\cite[Problem~1]{KruWer-0504166}.

Gill poses the following problem
in~\cite[Problem~26.B]{KruWer-0504166}: is there any Bell inequality
that holds for all quantum states, other than the inequalities
representing nonnegativity of probabilities?
Theorems~\ref{thrm:hypermetric-bell-facet-chsh} and
\ref{thrm:pure-cw-bell-facet-chsh} give a partial answer to this
problem.
If a Bell inequality $\vct{a}^\trans\vct{q}\le0$ includes the CHSH
inequality, then the Bell inequality $\vct{a}^\trans\vct{q}\le0$ is
necessarily violated in any quantum states violating the CHSH
inequality.

Further investigation of inclusion relation and families of Bell
inequalities may be useful to understand the structure of Bell
inequalities such as the answer to Gill's problem.

\section*{References}

\bibliography{bell}

\end{document}